\documentstyle[aps,twocolumn,psfig]{revtex}
\begin{document}
\title{On the Analyticity of Solutions in the Dynamical Mean-Field Theory}
\author{Th.~Pruschke$^{a,}$ \cite{apruschke}, W.~Metzner$^{b}$, and D.~Vollhardt$^{c}$}
\address{$^a$Institut f\"ur Theoretische Physik, Universit\"at Regensburg,
93040 Regensburg, Germany\\ $^{b}$Institut f\"{u}r Theoretische
Physik C, Technische Hochschule Aachen, 52056 Aachen, Germany\\
$^{c}$ Theoretische Physik III, Elektronische Korrelationen und
Magnetismus, Universit\"{a}t Augsburg, 86135 Augsburg, Germany}
\maketitle

\begin{abstract}
The unphysical solutions of the periodic Anderson model obtained
by H.~Keiter and T.~Leuders [Europhys.\ Lett., {\bf 49}, 801
(2000)] in dynamical mean-field theory (DMFT) are shown to result
from the author's restricted choice of the functional form of the
solution, leading to a violation of the analytic properties of the
exact solution. By contrast, iterative solutions of the
self-consistency condition within the DMFT obtained by techniques
which preserve the correct analytic properties of the exact
solution (e.g., quantum Monte-Carlo simulations or the numerical
renormalization group) always lead to physical solutions.
\end{abstract}

\date{\today}

In a recent paper, Keiter and Leuders \cite{keiter} discussed the structure
and the solutions of the self-consistency condition of the dynamical
mean-field theory (DMFT) for the periodic Anderson model at $U=\infty $.
They raised two independent points: (i), they showed analytically that on
certain assumptions for the structure of the solution the self-consistency
condition can have multiple solutions, which merge in the upper half of the
complex plane, and (ii), they argued that calculations of $1/d$ corrections (%
$d$: spatial dimension of the lattice) to the self-consistency condition of
the DMFT are problematic, since the limit $d\rightarrow \infty $ does not
commute with the limit $n\rightarrow \infty $ of the length $n$ of loops
embedded in the lattice.

In the following both points will be addressed. The main part of the paper
will be concerned with a detailed analysis of the first issue. In
particular, we will show that the conclusions of Keiter and Leuders \cite%
{keiter} relating to this point are unwarranted, i.e., are due to the
deficiency of the method used by these authors to solve (or simulate the
solution of) the single-impurity Anderson model (SIAM).

\subsection{Structure of the solution of the self-consistency condition}

Multiple solutions of a nonlinear set of equations are not unusual. In
particular, most multiple solutions of nonlinear equations for propagators
are unphysical, i.e., have poles in the upper half-plane or lead to an
incorrect asymptotic behavior (or both). In this respect the findings by
Keiter and Leuders \cite{keiter} that under certain conditions there exists
a branch point in the upper half-plane where an unphysical and the physical
solution to the self-consistency condition {\em merge}, is disturbing. If
true in general it would raise doubts about the uniqueness of numerically
obtained iterative solutions\/ to the self-consistency condition, and would
make the physical relevance of results obtained within the DMFT questionable.

Here we will show that, in fact, the self-consistency condition for the
periodic Anderson model (PAM) \cite{foot1} {\em always leads to a unique
physical result unless the functional space for the solutions is
artificially restricted} as done in ref.\cite{keiter}. In standard notation
the PAM reads
\begin{equation}
H_{{\rm PAM}}=%
\begin{array}[t]{l}
\displaystyle\sum\limits_{{\bf k}\sigma }\epsilon _{{\bf k}}c_{{\bf k}\sigma
}^{\dagger }c_{{\bf k}\sigma }^{\phantom{\dagger}} \\[5mm]
\displaystyle+\sum\limits_{i\sigma }\left( \epsilon _{f}+\frac{U}{2}f_{i\bar{%
\sigma}}^{\dagger }f_{i\bar{\sigma}}^{\phantom{\dagger}}\right) f_{i\sigma
}^{\dagger }f_{i\sigma }^{\phantom{\dagger}} \\[5mm]
\displaystyle+V\sum\limits_{i\sigma }\left( c_{i\sigma }^{\dagger
}f_{i\sigma }^{\phantom{\dagger}}+\mbox{h.c.}\right) .%
\end{array}
\label{PAM}
\end{equation}%
Within the DMFT, the nontrivial part of the partition function is
represented by an effective local action \cite{gkrev}
\begin{equation}
S_{{\rm eff}}=%
\begin{array}[t]{l}
\displaystyle-\sum\limits_{\sigma }\int d\tau d\tau ^{\prime }f_{\sigma
}^{\dagger }(\tau )\tilde{{\cal G}}^{-1}(\tau -\tau ^{\prime })f_{\sigma }^{%
\phantom{\dagger}}(\tau ^{\prime }) \\[5mm]
\displaystyle+U\int d\tau f_{\uparrow }^{\dagger }(\tau )f_{\uparrow }^{%
\phantom{\dagger}}(\tau )f_{\downarrow }^{\dagger }(\tau )f_{\downarrow }^{%
\phantom{\dagger}}(\tau )%
\end{array}
\label{Seff}
\end{equation}%
where \cite{footnote0}
\begin{equation}
\tilde{{\cal G}}(z)=\left[ z-\epsilon _{f}-V^{2}{\cal G}(z)\right] ^{-1}
\label{equ:0}
\end{equation}
and the auxiliary propagator ${\cal G}(z)$ describes an artificial system
where the $f$-states at the site under consideration have been removed. Note
that the action (\ref{Seff}) defines an effective SIAM with band-electron
Green function ${\cal G}(z)$ which has to be determined self-consistently.
For the PAM, (\ref{PAM}), Keiter and Leuders derived the following
expression \cite{keiter}:
\begin{equation}
{\cal G}(z)=\frac{1}{N}\sum\limits_{{\bf k}}\frac{g_{{\bf k}}(z)}{1-g_{{\bf k%
}}(z)T(z)+T(z){\cal G}(z)}.  \label{equ:1}
\end{equation}%
Here $g_{{\bf k}}(z)=1/(z-\epsilon _{{\bf k}})$ is the propagator of the
non-interacting conduction band system and $T(z)=V^{2}F(z)$, with $F(z)$ as
the local $f$-electron Green function. The latter relation implies that $%
T(z) $ cannot be chosen at liberty. Eq. (\ref{equ:1}) is not sufficient to
solve the problem, since $F(z)$ is unknown. The quantity $F(z)$ has to be
determined from an effective SIAM, where the propagator of the host
electrons is given by ${\cal G}(z)$.
It should be noted that the relation (\ref{equ:1})
is not only valid for $U=\infty$, but for {\em all} values of $U$, including
$U=0$. Moreover, it is important to mention that in Ref.\ \cite{keiter} the
artificial host propagator ${\cal G}(z)$ appearing in (\ref{equ:0}) is denoted
by $G(z)$, while conventionally, and also in this paper, $G(z)$ refers to the
propagator of the {\em physical band states}.

Let us briefly comment on the validity of Equ.\ (\ref{equ:1}) for $U=0$. At first
glance it seems rather strange that the solution to a non interacting system
should be given by a complicated, non-linear self-consistency condition. The
reason, however, is quite simple. First, Equ.\ (\ref{equ:1}) is set up in {\em
real space}, while the solution to (\ref{PAM}) for $U=0$ is naturally obtained
in reciprocal space. Second, as mentioned before, the quantity ${\cal G}(z)$ is
not a physical propagator but an artificial auxiliary quantity. It is perfectly
valid, although not necessarily reasonable, to rewrite the solution of the PAM
(\ref{PAM}) in terms of this auxiliary local propagator even for $U=0$, thus
obtaining the complicated expression (\ref{equ:1}). For the {\em physical}
propagator $G(z)$, on the other hand, one eventually finds, inserting
$T(z)=V^2\left[z-\epsilon_f-V^2{\cal G}(z)\right]^{-1}$, a simple and sensible
solution (cf.\ Equ.\ (\ref{equ:4}) below). It must be noted that for
$U=0$ this procedure is exact independent of dimension provided
the hybridization in (\ref{PAM}) is local. The DMFT or limit $d=\infty$ enters
at a different stage, namely through the locality of $T(z)$ {\em even in the presence
of the two-particle interaction $U$}.

To examine the analytic properties of (\ref{equ:1}) under the
condition that $T(z)$ is given by the solution to an effective SIAM
some simple algebraic transformations have to be performed. Let us assume
that $T(z)$ and ${\cal G}(z)$ are solutions to (\ref{equ:1}). 
Then we may rewrite (\ref{equ:1}) as
\begin{equation}
{\cal G}(z)=\frac{1}{1+T(z){\cal G}(z)}\frac{1}{N}\sum\limits_{{\bf k}}\frac{%
1}{z-\epsilon _{{\bf k}}-\frac{T(z)}{1+T(z){\cal G}(z)}}.  \label{equ:2}
\end{equation}%
For the SIAM, the host-electron propagator of the interacting system is
obtained from the host propagator of the system without $f$-states by
\begin{equation}
G(z)={\cal G}(z)+{\cal G}(z)T(z){\cal G}(z).  \label{equ:3}
\end{equation}%
We can rewrite (\ref{equ:1}) in terms of the physical host-electron
propagator as
\begin{equation}
G(z)=\frac{1}{N}\sum\limits_{{\bf k}}\frac{1}{z-\epsilon _{{\bf k}}-\frac{%
T(z)}{1+T(z){\cal G}(z)}},  \label{equ:4}
\end{equation}%
and thereby identify the self-energy
\begin{equation}
\Sigma ^{(c)}(z)=\frac{T(z)}{1+T(z){\cal G}(z)}  \label{equ:5}
\end{equation}%
of the host electrons. Note that for $U=0$ the self energy (\ref{equ:5}) reduces to the
well-known expression $\Sigma ^{(c)}(z)=V^2/[z-\epsilon_f]$. Since the ${\bf k}$-dependence in (\ref{equ:4}) is
only due to $\epsilon _{{\bf k}}$, the ${\bf k}$-sum can be converted to an
energy integral
\begin{equation}
\frac{1}{N}\sum\limits_{{\bf k}}\ldots \;\rightarrow \;\int d\epsilon \rho
^{(0)}(\epsilon )\ldots   \label{DOS}
\end{equation}%
with a positive-definite density of states (DOS) $\rho ^{(0)}(\epsilon )$.
The relation (\ref{equ:4}) is simply a Hilbert transform; therefore one can
make use of its analytical properties to show \cite{maier}
\begin{equation}
G(z)=\frac{1}{z-\Sigma ^{(c)}(z)-\Gamma (z-\Sigma ^{(c)}(z))},  \label{equ:6}
\end{equation}%
with a Herglotz-analytic function $\Gamma (\zeta )$, i.e.
$${\rm sign}\left( \Im
m\Gamma (\zeta )\right) =-{\rm sign}\left( \Im m\zeta \right)
$$
and $\Gamma
(\zeta )\sim 1/\zeta $ for $|\zeta |\rightarrow \infty $. Furthermore, $%
\Gamma (\zeta )$ is, in principle, a known function which is determined by
the structure of the DOS alone. Finally, ${\cal G}(z)$ can be obtained from
the definition of $G(z)^{-1}={\cal G}(z)^{-1}-\Sigma ^{(c)}(z)$ as
\begin{equation}
{\cal G}(z)^{-1}=G(z)^{-1}+\Sigma ^{(c)}(z)=z-\Gamma (z-\frac{T(z)}{1+T(z)%
{\cal G}(z)})\;\;.  \label{equ:7}
\end{equation}%
It should be stressed that (\ref{equ:7}) is not merely a nonlinear {\em %
equation}\/ determining ${\cal G}(z)$ but, rather, a {\em functional equation%
}, since $T(z)$ is a functional of ${\cal G}(z)$. The functional
dependence is determined by the solution of a nontrivial problem,
namely the SIAM. It is thus not sufficient merely to study the
possible roots of the nonlinear equation (\ref{equ:7}), but one
has to discuss the structure of this equation under the constraint
that $T(z)$ must obey the basic properties of the solution to the
SIAM (\ref{Seff}). Since there does not exist an analytic solution
to the SIAM in closed form this constraint can be incorporated
currently only within an iterative scheme, even if $\Gamma (\zeta
)$ is known in closed form as, for example, for the Bethe lattice
or the toy DOS used in \cite{keiter}.

Let us therefore assume that we knew the exact solution of the SIAM, and
study the consequences of an iterative procedure to solve the
self-consistency condition. Starting in the $n$-th step of this
iterative procedure with a physical ${\cal G}^{(n)}(z)$ with $%
{\rm sign}\left( \Im m{\cal G}^{(n)}(z)\right) =-{\rm sign}\left( \Im mz\right) $
and ${\cal G}^{(n)}(z)\sim 1/z$ ($|z|\rightarrow \infty $), the
corresponding propagator of the $%
f$-states, $F^{(n)}(z)$, and hence $T^{(n)}(z)$, will have the same
properties, i.e., the right-hand side of (\ref{equ:7}) will definitely have the correct asymptotics, because $T^{(n)}(z)%
{\cal G}^{(n)}(z)\sim 1/z^{2}$ and thus
\[
\frac{T^{(n)}(z)}{1+T^{(n)}(z){\cal G}^{(n)}(z)}\sim 1/z
\]%
as $|z|\rightarrow \infty $. More importantly, however, the exact
solution to the effective SIAM (\ref{Seff}) has to satisfy the
condition \cite{footnote2}
\begin{equation}
{\rm sign}\left[ \Im m\left( T^{(n)}(z)^{-1}+{\cal G}^{(n)}(z)\right) \right] =+{\rm sign%
}\left( \Im mz\right)   \label{unitarity}
\end{equation}%
which is necessary to ensure that $\Sigma ^{(c)}(z)$,
(\ref{equ:5}), has the correct analytic properties, i.e., \ ${\rm
sign}\left( \Im m\Sigma ^{(c)}(z)\right) =-{\rm sign}\left( \Im
mz\right) $. Under those conditions, the iterative procedure
always stays within the subspace of physical
solutions. Here it must be stressed once more that one is not free to {\em %
choose} $T^{(n)}(z)=V^{2}F^{(n)}(z),$ or even choose a particular functional form for $%
T^{(n)}(z),$ since this form is entirely and uniquely determined by the solution
of the effective SIAM with a fixed host Green function ${\cal G}^{(n)}(z)$.
Especially condition (\ref{unitarity}) turns out to be very susceptible to
approximations and is easily violated. Such a violation would manifest
itself in an indefinite sign of $\Im m\Sigma (z)$ and would thus lead to the
crossing of a branch point into the upper half plane. Note also that
in the course of the iteration, the quanitities entering on the
right-hand side of
(\ref{equ:7}) are those of the $n$-th step, producing the host Green
function
${\cal G}^{(n+1)}(z)$ for the $n+1$-th step and, unless
self-consistency has been reached, ${\cal G}^{(n+1)}(z)\ne{\cal
  G}^{(n)}(z)$ in general \cite{footnote3}.

\begin{figure}
\begin{center}\mbox{}
\psfig{figure=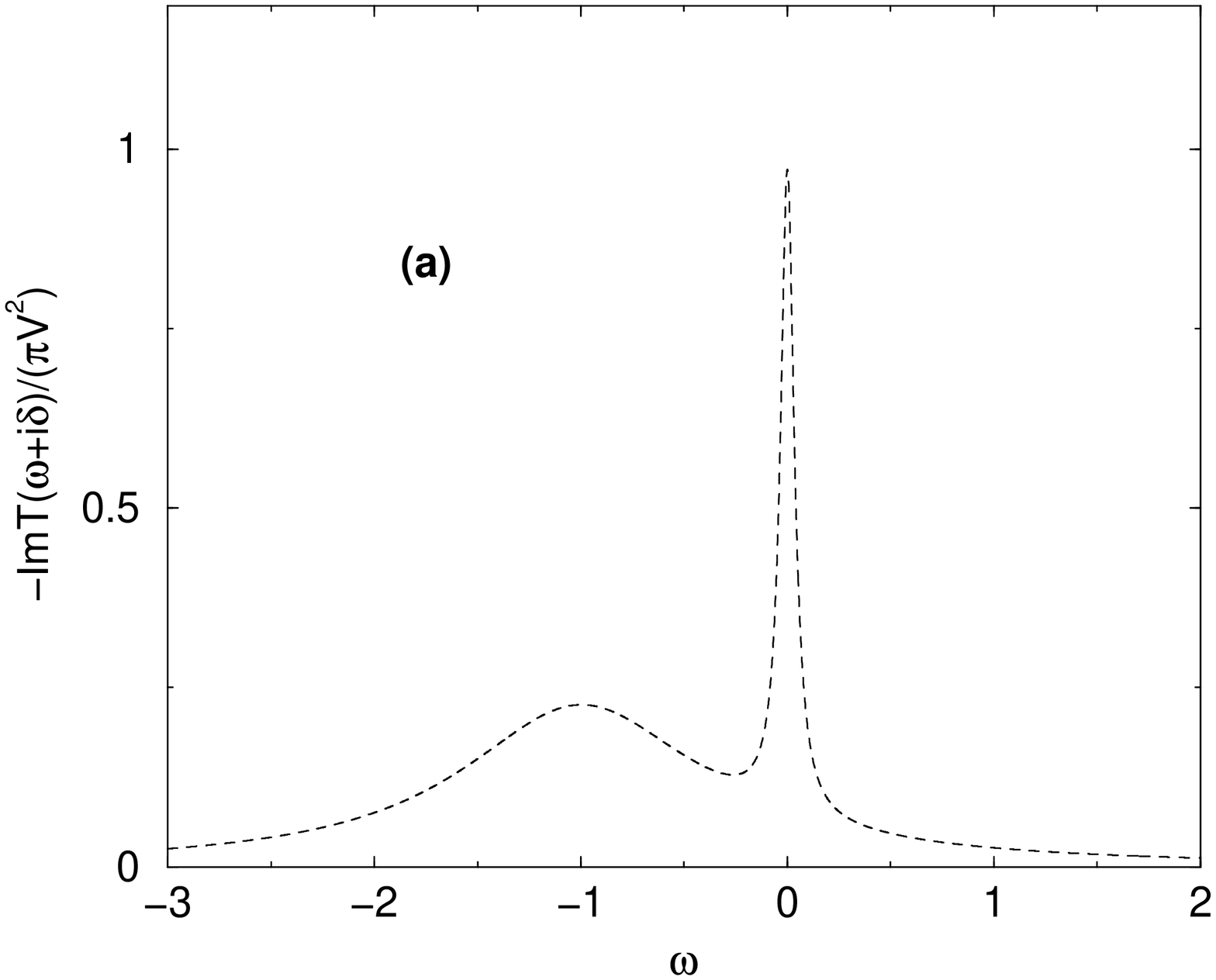,angle=-90,width=0.375\textwidth}
\end{center}
\begin{center}\mbox{}
\psfig{figure=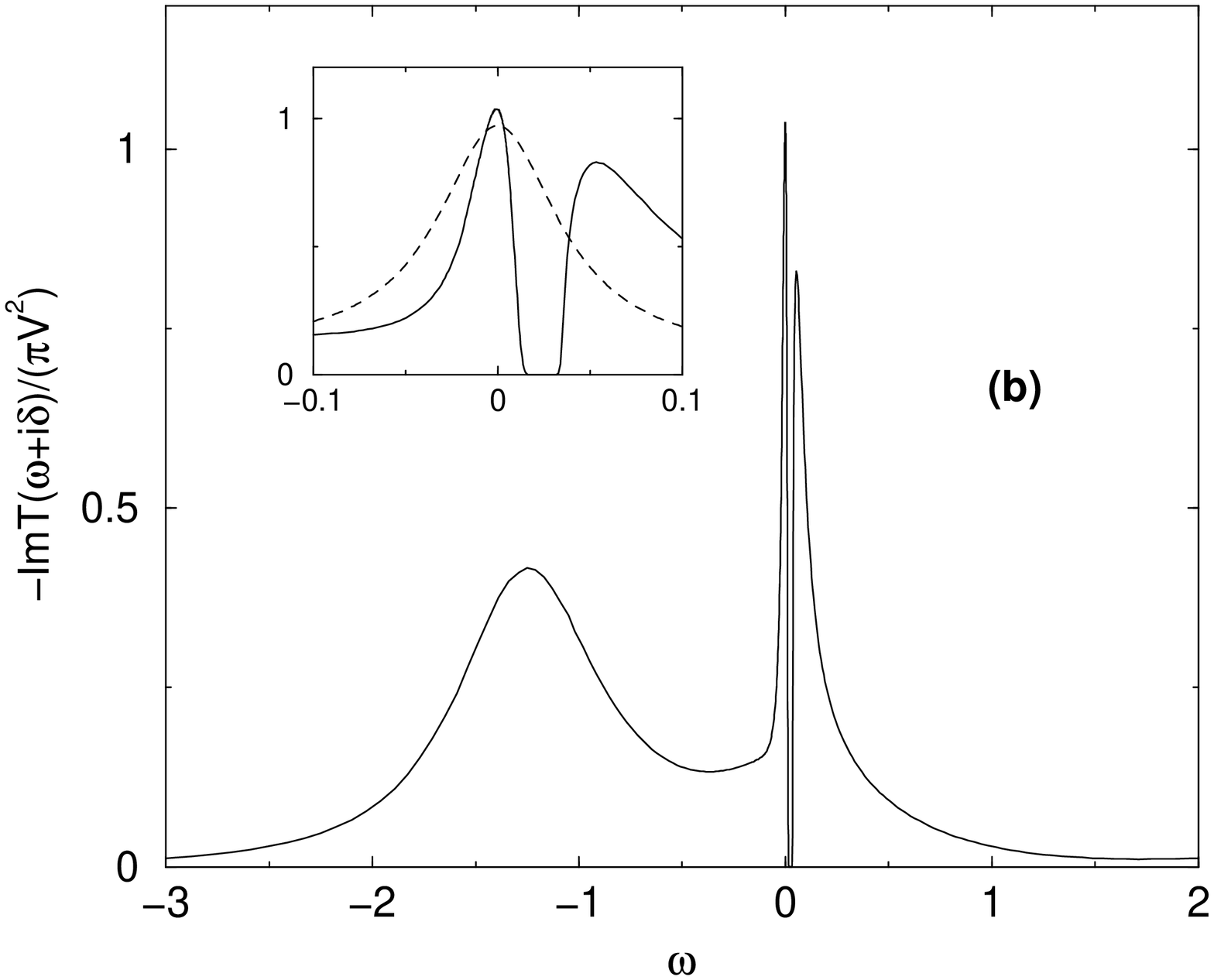,angle=-90,width=0.375\textwidth}
\end{center}
\caption[]{Imaginary part of $T(\omega+i\delta)$ for the PAM. (a)
result according to (\ref{leuders}) as used in ref. \cite{keiter};
(b) result from a full DMFT calculation with the NRG \cite{pbj}. In the inset
in (b) the two curves are compared for energies close to the Fermi
energy. } \label{fig:1}
\end{figure}

Hence we find that the branch point in the upper half-plane observed by
Keiter and Leuders\cite{keiter} is merely due to the author's restricted
choice of the functional form of $T(z)$ which leads to a violation of (\ref%
{unitarity}). Namely, motivated by results for the standard SIAM these
authors {\em chose} for $T(z)$ the following ansatz \cite{leudersdiss}
\begin{equation}
T(z)=V^{2}\left( \frac{a_{\epsilon _{f}}}{\omega -\epsilon _{f}+i\Gamma
_{\epsilon _{f}}}+\frac{a_{{\rm K}}}{\omega +i\Gamma _{{\rm K}}}\right) \;\;.
\label{leuders}
\end{equation}%
This form represents the strong-coupling regime of the SIAM, i.e.\
$-\epsilon_f$, $\epsilon_f+U\gg\pi\rho^{(0)}(0)V^2$. 
The original single-particle level at
$\epsilon _{f}$ is broadened by the coupling to the conduction
bath. In addition, the sharp Abrikosov-Suhl resonance which
characterizes the virtual bound state generated by spin-flip
scattering appears at the Fermi energy. A typical result will have
the form shown in Fig.\ \ref{fig:1}a. In Fig.\ \ref{fig:1} we assumed
a Gaussian free DOS $\rho^{(0)}(\epsilon)=\exp(-\epsilon^{2})/\sqrt{\pi }$ 
and set $V^{2}=0.2$, $\epsilon _{f}=-1$. Let us compare this result
with a fully self-consistent DMFT calculation for the PAM with the
same parameters \cite{pbj}. The result is shown in Fig.\ \ref{fig:1}b
(full line). It is obvious from the inset to Fig.\ \ref{fig:1}b that
the true solution of the PAM does not have the simple structure given
by (\ref{leuders}) assumed in ref.\ \cite{keiter}. In particular, one
does not observe a single peak at the Fermi energy, but two peaks
separated by a gap.

Consequently, the conclusion following equ.\ (10) in
ref.\cite{keiter} is unwarranted: the observed breakdown, i.e.\
the appearance of a branch point in the upper half plane, does not
result from the fact that in this regime physically interpretable
single-valued solutions cannot be found, but merely from the
deficiency of the method used to solve (or simulate the solutions
of) the SIAM. In particular, one of the authors \cite{Pruschke}
and F.~Anders \cite{anders}
observed that approximations based on the resolvent perturbation theory \cite%
{kk}, like the non-crossing approximation (NCA) or improved
versions of it, always lead to spectra similar to (\ref{leuders})
and are not able to reproduce more complicated features like the
double-peak structure at the Fermi energy appearing in Fig.\
\ref{fig:1}b. It is therefore not surprising that the application
of these techniques in
solving the DMFT for the PAM leads to the breakdown described in ref.\cite%
{keiter}: it is a problem generated by the method used to solve
the self-consistency condition and not of the self-consistency
condition itself.

Our arguments do not imply that the self-consistency condition
(\ref{equ:7}) must necessarily have only one unique solution for a
given set of model parameters. Such a conclusion would require a
more detailed knowledge of the solutions of the SIAM (especially,
whether there exists a one-to-one correspondence between ${\cal
G}(z)$ and $F(z)$), and of the extent of the basin of attraction
for a particular solution. Since there does not exist an
exact analytical solution to the SIAM for an arbitrary host Green function $%
{\cal G}(z)$, statements about these properties of the self-consistency
condition are out of reach. Indeed for the Hubbard model such a situation
(namely, two physically valid, coexisting solutions) occurs close to the
Mott-Hubbard metal-insulator transition \cite{gkrev}.

\subsection{$1/d$ corrections to the DMFT}

In DMFT a fermionic lattice model is mapped onto a SIAM, complemented by a
self-consistency condition \cite{gkrev}. This mapping is rigorous in $%
d=\infty $, a fact which is not disputed in ref.\cite{keiter}.

The possibilities and limitations of systematic $1/d$-expansions around $%
d=\infty $ have already been explored by a number of authors
\cite{florian,vlaming92,vlaming2,pvd,schiller,jarrell,jv01}. Whether
$1/d$-corrections to DMFT are
able to improve the theory depends on the quantity under investigation, on
the type of $1/d$-expansion, and also on the physics one would like to
describe. For example, while $1/d$-corrections include non-local short-range
correlations they are unable to capture certain long-range correlations.
Furthermore, $1/d$-corrections, when truncated at finite order, may violate
basic analytical properties of dynamical quantities \cite{pvd}. All this is
not unexpected. Indeed, analogous problems arise in $1/d$-expansions around
the Weiss mean-field theory for the Heisenberg model and the coherent
potential approximation for the Anderson disorder model \cite{vlaming92}.
We do not see why the peculiarities of large loops in the large $d$
limit emphasized by Keiter and Leuders \cite{keiter} should make
expansions around DMFT particularly difficult or ambiguous.
Since even the Green function of the {\em non-interacting} system can be
expanded in loops, it is clear that these peculiarities are in one-to-one
correspondence with the special behavior of the {\em bare}\/ DOS in the
limit $d\rightarrow \infty $ (e.g., the absence of van Hove singularities)
\cite{mueha89,bethe}.
Potential problems with the $1/d$-expansion of the bare DOS can be avoided
by inserting the actual DOS of the finite-dimensional system into the DMFT
equations. Then only corrections to the {\em local approximation} of
irreducible vertices underlying the DMFT need to be considered, as has been
done already by several authors \cite{florian,jarrell,metzner89,maier00}.

\bigskip

In summary, we showed that the iterative solution of the self-consistency
condition within the DMFT always lead to solutions that stay in the physical
subspace. Therefore, the unphysical branch cuts observed by Keiter and Leuders%
\cite{keiter} are due to the method used by these authors to solve
the SIAM which violates fundamental analytic properties of the
exact solution. Similar violations are caused by the NCA and other
approximations based on the resolvent perturbation theory
\cite{kk}. In fact, this also appears to be the reason why it was
not possible up to now to obtain meaningful results for the PAM
using NCA and related methods, while techniques which preserve the
correct analytic properties of the exact solution (e.g., iterated
perturbation theory, quantum Monte-Carlo simulations or the
numerical renormalization group) do not encounter similar
difficulties.

\bigskip

We thank F.~Anders, V.~Jani\v{s}, M.~Jarrell, H.~Keiter and D.~Otto for useful communication.


\begin{references}
\bibitem{foot1} Here we shall focus on the PAM in order to make direct
contact with the discussion in ref.\cite{keiter}. However, the arguments
presented are valid also for general one-band models.
\bibitem{footnote0} The appearance of a momentum independent hybridization $V
$ in (\ref{PAM}) implies that the representation of $\tilde{{\cal G}}(z)$ is
not unique. Therefore the form of $\tilde{{\cal G}}(z)$ used in ref.\ \cite%
{keiter} and here differs from the one used in ref.\ \cite{gkrev},
but can be derived from it using simple algebra.
\bibitem{footnote2} This relation follows quite simply from the fact that $%
T(z)=V^{2}F(z)$, where the local propagator $F(z)$ has the form
$F(z)=\left[ z-\epsilon _{f}-V^{2}{\cal G}(z)-\Sigma
^{(f)}(z)\right] ^{-1}$ with a Herglotz-analytic self-energy
$\Sigma ^{(f)}(z)$.
\bibitem{footnote3} One consequence is that, for an iterative
  solution, (\ref{equ:2}) is not a valid reformulation of
  (\ref{equ:7}), because it only holds when self-consistency has
  been reached.
\bibitem{bethe} Indeed, for the Bethe lattice, where self-avoiding loops do
not exist, the large-loop peculiarities discussed in Ref. \cite{keiter} and
the unphysical band tails typical for hypercubic lattices in $d\rightarrow
\infty $ are absent. Nevertheless the DMFT and (\ref{equ:6}) is well-defined
also for a Bethe lattice.
\end{references}
\end{document}